\documentclass[]{ceurart}

\usepackage{listings}
\usepackage[most]{tcolorbox}

\newcommand{\bslabel}[1]{\par\noindent\textbf{#1}}

\begin{document}

\copyrightyear{2026}
\copyrightclause{Copyright for this paper by its authors.
  Use permitted under Creative Commons License Attribution 4.0
  International (CC BY 4.0).}

\conference{CLEF 2026 Working Notes, 21 -- 24 September 2026, Jena, Germany}

\title{Claim2Source at CheckThat! 2026: Improving Multilingual Scientific Claim–Source Retrieval with Verification-based Re-Ranking}

\title[mode=sub]{Notebook for the CheckThat! Lab at CLEF 2026}

\author[1]{Tobias Schreieder}[%
orcid=0009-0000-8268-4204,
email=tobias.schreieder@tu-dresden.de,
]
\cormark[1]
\address[1]{TU Dresden \& ScaDS.AI Dresden/Leipzig, Dresden, Germany}

\author[2]{Harsh Khandelwal}[%
orcid=0009-0004-0237-5735,
email=harsh.khandelwal@fau.de,
]
\address[2]{Friedrich-Alexander University of Erlangen–Nuremberg, Erlangen, Germany}

\author[1]{Yu-Ling Zhong}[%
orcid=0009-0007-2361-4944,
email=yu-ling.zhong@mailbox.tu-dresden.de,
]

\author[1]{Michael Färber}[%
orcid=0000-0001-5458-8645,
email=michael.faerber@tu-dresden.de,
]

\cortext[1]{Corresponding author.}

\begin{abstract}
Multilingual scientific claim--source retrieval aims to identify the scientific publication supporting a claim shared on social media. This task is challenging because claims often differ from source publications in terms of language, wording, and level of detail, which weakens the connection between claims and their underlying evidence. In this paper, we present our approach for the CheckThat! 2026 Lab Task 1: \textit{Source Retrieval for Scientific Web Claims}. We propose a multi-stage retrieval framework for multilingual scientific claim--source retrieval that combines structured claim and source representations with progressive candidate refinement. To address multilingual retrieval challenges, the framework employs bilingual claim representations, metadata-enhanced source representations, and language-specific adaptation of dense retrieval models. Building on this setup, a first-stage retriever generates an initial pool of candidate sources, after which similarity-based re-ranking improves the ranking of highly relevant sources and verification-based re-ranking identifies the candidate source that best supports the claim using verification signals. Our approach achieves an average MRR@5 score of 0.7628 across English, German, and French claims, ranking first on the CheckThat! 2026 leaderboard.
\end{abstract}

\begin{keywords}
  Fact Verification \sep
  Scientific Claim-Source Retrieval \sep
  Multilinguality \sep
  Re-Ranking \sep
  Large Language Model
\end{keywords}

\maketitle

\section{Introduction}
\label{sec:introduction}

Social media platforms such as X (formerly Twitter) have become important channels for communicating and discussing scientific findings, particularly in fast-moving domains such as health and biomedicine~\cite{Guenther2023TwitterScienceCommunication}. Scientific claims shared online often discuss research findings without explicitly attributing them to their original scientific source. Identifying this source is challenging because claims often contain only partial descriptions of research findings and may rephrase or simplify the original evidence, creating a semantic mismatch between claims and their underlying publications~\cite{Schreieder2025Claim2Source,Schofield025DSGT}. This problem becomes even more difficult when claims and source documents are written in different languages, requiring retrieval systems to align semantically equivalent content across heterogeneous representations~\cite{Zhang2023MultilingualDenseRetrieval,Litschko2022CrossLingual}.

Attributing claims to their supporting evidence is an important component of automated fact verification systems, which aim to distinguish factual information from misinformation shared online. This problem is studied in the CheckThat! 2026 Lab, which focuses on advancing multilingual fact-checking~\cite{Struss2026CheckThat2026}, through Task 1: \textit{Source Retrieval for Scientific Web Claims}~\cite{Schellhammer2026CheckThatTask1}. The task was introduced in 2025~\cite{Alam2026CheckThat2025} and is extended in 2026 toward a multilingual setting with English, German, and French claims~\cite{Schellhammer2026CheckThatTask1}. Throughout this paper, we refer to this task as multilingual scientific claim--source retrieval. While previous systems mainly focused on improving retrieval performance through multi-stage retrieval and re-ranking pipelines~\cite{Sager2025DeepRetrieval,Staudinger025ATOM}, cross-lingual retrieval introduces additional challenges, as scientific literature remains predominantly available in English.

Building on recent findings that multi-stage retrieval and re-ranking improve scientific claim--source retrieval performance, we present a retrieval framework for multilingual settings, designed to address the challenges arising when claims and source publications are expressed in different languages. To handle multilingual claims, we employ bilingual claim representations and investigate language-specific adaptation of dense retrieval models. We further propose a novel three-stage retrieval architecture consisting of dense candidate generation, similarity-based re-ranking, and a verification-based re-ranking stage for progressive candidate refinement. In contrast to conventional retrieval pipelines that primarily rely on similarity estimation, the final stage directly incorporates verification signals into the retrieval process to improve source selection. Our contributions are summarized as follows:

\begin{enumerate}
    \item We introduce bilingual claim representations that combine original and translated claim formulations, preserving language-specific information while reducing language mismatch.
    
    \item We fine-tune GritLM-7B on multilingual training data spanning English, German, and French to improve retrieval performance across different claim languages.

    \item We develop metadata-enhanced source representations that combine publication content (title and abstract) with contextual metadata (authors and venue).
    
    \item We propose a novel three-stage retrieval framework comprising first-stage retrieval, similarity-based re-ranking, and a verification-based re-ranking stage using a large language model (LLM) for source selection that directly incorporates verification signals into ranking decisions.
\end{enumerate}

\section{Related Work}
\label{sec:related_work}

\bslabel{Scientific Claim--Source Retrieval.}
Scientific claim--source retrieval aims to identify the scientific publication underlying a claim expressed in social media or online discourse. In contrast to conventional evidence retrieval tasks, scientific claims frequently reference publications only implicitly and often differ substantially from scientific texts in language and style. This problem has been studied in the context of the CheckThat! Lab, which has established scientific claim--source retrieval as a benchmark task for evaluating retrieval approaches in realistic multilingual settings~\cite{Alam2026CheckThat2025,Struss2026CheckThat2026}. Recent systems have explored multi-stage retrieval pipelines to bridge the gap between informal social media claims and formal scientific publications. Sager et al.~\cite{Sager2025DeepRetrieval} demonstrated that combining sparse and dense retrieval with LLM-based re-ranking substantially improves retrieval performance. Staudinger et al.~\cite{Staudinger025ATOM} showed that listwise re-ranking consistently outperforms pointwise approaches. Schofield et al.~\cite{Schofield025DSGT} investigated data augmentation and claim reformulation strategies, showing that combining multiple claim representations can improve retrieval performance, whereas replacing original claims may remove important contextual information. The effectiveness of zero-shot style transfer methods for scientific claim--source retrieval was shown to depend strongly on the underlying retrieval model, with most sparse and hybrid models benefiting from more formal claim representations~\cite{Schreieder2025Claim2Source}. These findings indicate that scientific claim--source retrieval benefits from multi-stage architectures with candidate re-ranking and adaptation of claim and document representations, which motivate the design choices of our proposed framework.

\bslabel{Multilingual and Cross-Lingual Retrieval.}
Cross-lingual retrieval introduces additional challenges because claims and source documents may differ in language while still requiring alignment at the semantic level. Recent work has shown that dense multilingual representations can effectively align semantically similar queries and documents across different languages~\cite{Zhang2023MultilingualDenseRetrieval}. Lawrie et al.~\cite{Lawrie2023MultilingualIR} demonstrated that multilingual neural retrieval models can achieve performance close to translation-based retrieval pipelines while reducing indexing costs. Furthermore, Litschko et al.~\cite{Litschko2022CrossLingual} showed that cross-lingual retrieval can substantially benefit from in-domain fine-tuning of multilingual encoders. In the context of scientific retrieval, Valentini et al.~\cite{valentini-etal-2025-clirudit} found that multilingual dense representations and translation-based methods offer complementary strengths for cross-lingual retrieval, with translation particularly benefiting sparse retrieval models. Overall, these studies highlight that retrieval strategies depend strongly on the retrieval setting and underlying models, motivating the exploration of complementary approaches such as translation, bilingual representations, and language-specific adaptation.

\bslabel{Re-Ranking and Verification.}
Re-ranking has become a central component of scientific claim--source retrieval, as the best-performing approaches in the CheckThat! 2025 shared task predominantly relied on retrieval pipelines with re-ranking stages~\cite{Alam2026CheckThat2025}. Classical re-ranking approaches primarily employ semantic similarity estimation to refine candidate rankings~\cite{Staudinger025ATOM}. More recently, LLMs have been integrated into retrieval pipelines for candidate re-ranking. Existing work explored LLM-based source filtering techniques to improve the selection of retrieved candidates~\cite{Besrour2025SQuAI,Besrour2025RAGentA}. Beyond candidate ranking, verifier feedback has been used to optimize evidence retrieval by measuring how useful retrieved evidence is for downstream claim verification~\cite{Zhang2023FeedbackFactVerification}. Together, these developments underscore that verification signals can provide complementary information beyond semantic similarity. This motivates our use of verification-based re-ranking for refining highly ranked candidate sources.

\section{Dataset}
\label{sec:dataset}

The experiments are conducted on the scientific claim--source retrieval dataset used in CheckThat! 2026\footnote{\url{https://huggingface.co/datasets/sschellhammer/CT26_Task1_SourceRetrievalForScientificWebClaims}}~\cite{Struss2026CheckThat2026}. The dataset contains social media claims paired with their corresponding scientific source publications. Claims are available in English, German, and French, while sources are in English. Table~\ref{tab:dataset} summarizes the distribution across languages and data splits. The objective is to identify the scientific publication referenced by a claim from a collection of 10,000 candidate documents within a known-item retrieval setting. Compared to the previous edition, the multilingual setup increases task complexity because claims and source documents may differ not only in style but also in language, requiring retrieval systems to align semantically equivalent information across languages. We use the training split for model fine-tuning and the validation split for model selection and ablation studies, while the final system performance is evaluated on the official test set.

\begin{table}[t]
\centering
\caption{CheckThat! 2026 dataset statistics across train, validation, and test splits for English (EN), German (DE), and French (FR) claims. The retrieval corpus consists of 10,000 scientific publications used as candidate sources.}
\label{tab:dataset}
\setlength{\tabcolsep}{4pt}
\begin{tabular}{lccc}
\toprule
\textbf{Split} & \textbf{EN} & \textbf{DE} & \textbf{FR} \\
\midrule
Train      & 14,977 & 1,460 & 2,807 \\
Validation & 3,905  & 386   & 702 \\
Test       & 6,076  & 876   & 1,220 \\
\bottomrule
\end{tabular}
\end{table}

\section{Methodology}
\label{sec:methodology}

\begin{figure}[t]
    \centering
    \includegraphics[width=\linewidth]{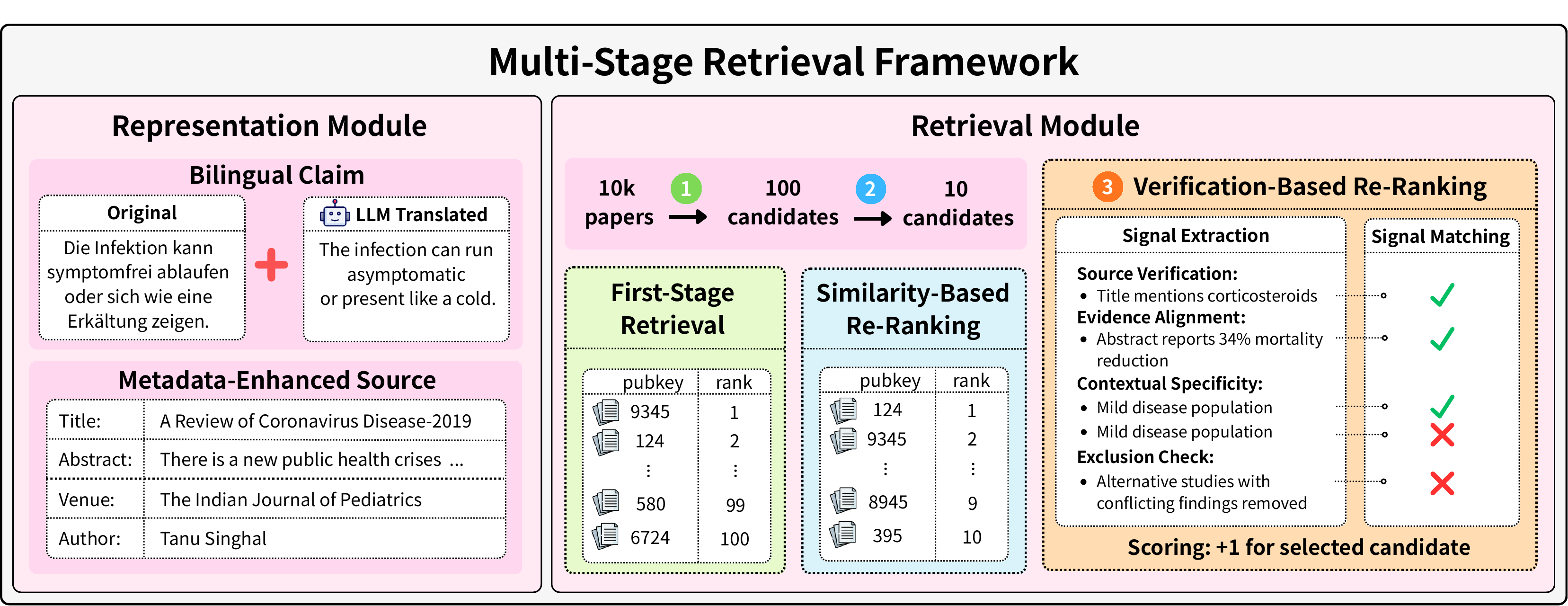}
    \caption{Multi-stage retrieval framework for multilingual scientific claim--source retrieval. The framework consists of a representation module and a retrieval module. The representation module constructs suitable representations for claims and sources: claims are represented as original–translated claim pairs, while sources are encoded using metadata-enhanced structured templates. The retrieval module progressively refines candidate selection through first-stage retrieval, similarity-based re-ranking, and verification-based re-ranking.}
    \label{fig:multi_stage_retrieval_framework}
\end{figure}

Our approach combines tailored claim and source representations with a multi-stage retrieval framework for multilingual scientific claim--source retrieval (see Figure~\ref{fig:multi_stage_retrieval_framework}). First, we construct claim and source representations designed to reduce language mismatch and expose additional source signals. Subsequently, candidate sources are progressively refined through a three-stage retrieval process consisting of first-stage retrieval, similarity-based re-ranking, and verification-based re-ranking.

\subsection{Claim and Source Representation}
\label{sec:claim_source_representation}

Social media claims and scientific source publications may differ substantially in language and level of detail, while correct source attribution often depends on signals beyond the publication content alone. To address this, we introduce bilingual claim representations and metadata-enhanced source representations that reduce language mismatch and incorporate additional contextual source signals.

\subsubsection{Bilingual Claim Representation}
\label{sec:multilingual_claims}

We translate German and French claims into English in a zero-shot setting using the open-source LLM \textit{Qwen/Qwen3-8B}. Rather than replacing the original claim formulation, we construct a bilingual representation by combining the original and translated versions of each claim. This preserves language-specific information while providing an English formulation better aligned with the predominantly English source collection. The translation prompt is provided in Figure~\ref{fig:translation_prompt} in Appendix.

\subsubsection{Metadata-Enhanced Source Representation}
\label{sec:metadata_sources}

Scientific claims often refer to publications indirectly through mentions of authors, institutions, or publication venues rather than explicit citations. To capture such signals, we augment source representations with structured publication metadata in addition to the textual publication content. Specifically, each source publication is represented using labeled fields for the \textit{title}, \textit{abstract}, \textit{authors}, and \textit{venue}. To reduce noise from long author lists, only the first and last three authors are included.

\subsection{Multi-Stage Retrieval Framework}
\label{sec:multistage_retrieval}

Correctly identifying the scientific source underlying a social media claim often requires progressive refinement, since the two may differ substantially in terms of how information is expressed and which information is covered. We therefore develop a three-stage retrieval framework consisting of first-stage retrieval for candidate generation, similarity-based re-ranking for semantic candidate refinement, and verification-based re-ranking for identifying the candidate source that best supports the claim. Multiple open-source models are considered at each stage, and the best-performing configuration is selected for the final CheckThat! 2026 submission.

\subsubsection{First-Stage Retrieval}
\label{sec:retrieval}

The first stage of our framework generates an initial ranking of candidate sources from a collection of 10,000 scientific publications. To capture different retrieval paradigms, we evaluate both sparse and dense retrieval models, while additionally investigating language-specific adaptation of the best-performing dense retrieval model across different claim languages.

\bslabel{BM25.} \textit{BM25} is a sparse retrieval model based on term frequency and inverse document frequency \cite{Robertson1994BM25,Robertson1994Okapi}. Its retrieval behavior depends strongly on overlap between claim and source terminology.

\bslabel{GTR.} \textit{gtr-t5-xl} is a dense retrieval model with 11B parameters based on the T5 architecture and trained on large-scale retrieval tasks~\cite{Ni2022GTR}. The model is designed to capture semantic similarity.

\bslabel{E5.} \textit{intfloat/e5-large-v2} is a dense retrieval model with 335M parameters trained using contrastive learning on large-scale text-pair datasets for semantic retrieval~\cite{Wang2024E5}.

\bslabel{GritLM.} \textit{GritLM/GritLM-7B} is a dense retrieval model with 7B parameters based on a decoder-only transformer architecture. It is trained with generative representational instruction tuning (GRIT), jointly optimizing generation and retrieval objectives within a unified model~\cite{Muennighoff2025GritLM}. The resulting model achieves strong retrieval performance in zero-shot settings while retaining generative capabilities.

\bslabel{GritLM-F.} We further construct a fine-tuned variant of GritLM, using low-rank adaptation (LoRA) on the multilingual training split of the dataset. Fine-tuning is performed using claims and their corresponding source publications across all available languages, while additionally incorporating hard negative source samples in a contrastive retrieval training setup. This adaptation aims to better align retrieval representations with the multilingual characteristics of the task. Detailed training settings and hyperparameters are reported in Table~\ref{tab:gritlm_finetuning} in Appendix.

\subsubsection{Similarity-based Re-Ranking}
\label{sec:similarity_reranking}

The second stage of the proposed framework involves refining the initial source ranking using similarity-based re-rankers. Starting from the top-100 candidate sources retrieved by the selected first-stage retrieval model, re-ranking models score the relevance of each claim--source pair based on their textual representations and generate an updated ranking.

\bslabel{Nemotron.} The \textit{nvidia/llama-nemotron-rerank-vl-1b-v2} model is a pointwise cross-encoder re-ranker with approximately 1.7B parameters. Cross-encoders encode a claim and a source document together as a single input sequence, then score the relevance of the pair individually.

\bslabel{BGE.} Similar to Nemotron, \textit{BAAI/bge-reranker-v2-m3} is a pointwise cross-encoder re-ranker, but with a smaller model size of approximately 0.6B parameters and multilingual training objectives designed for cross-lingual retrieval settings~\cite{Chen2024M3}.

\bslabel{Jina.} The \textit{jinaai/jina-reranker-v3} model is a multilingual listwise re-ranker with approximately 0.6B parameters~\cite{Wang2025Jina}. Unlike pointwise approaches, listwise re-ranking considers the relative relevance of multiple candidate sources within a shared context to determine their ranking.

\bslabel{Qwen3.} The models \textit{Qwen/Qwen3-Reranker-0.6B} and \textit{Qwen/Qwen3-Reranker-8B} are pointwise generative re-rankers from the Qwen3 family~\cite{Zhang2025Qwen3}. In contrast to conventional cross-encoder re-rankers, Qwen3 uses a generative formulation for relevance estimation and enables analysis of the effects of model scale.

\bslabel{Qwen3-IT.} We additionally evaluate an instruction-tuned variant with a customized prompt designed for multilingual scientific claim--source retrieval. The adapted prompt incorporates task-specific guidance to investigate whether stronger instruction alignment improves re-ranking performance. The full prompt template is provided in Figure~\ref{fig:qwen_reranker_prompt} in Appendix.

\subsubsection{Verification-based Re-Ranking}
\label{sec:verification_reranking}

The final stage performs verification-based re-ranking to refine the highest-ranked candidate sources. While similarity-based approaches improve semantic matching between claims and source documents, semantically similar candidates may still differ in the evidence they provide and not necessarily support the claim. To address this limitation, we incorporate verification signals into multilingual scientific claim--source retrieval, following ideas from fact verification literature where evidence assessment is used to determine whether retrieved information supports a claim~\cite{Thorne2018FEVER}. The first two stages already reduce the candidate space to a small set of highly relevant sources, such that the correct source is frequently contained among the top-ranked documents. Verification-based re-ranking therefore focuses on identifying the best matching source and promoting it to the highest rank position.

Due to the computational cost of LLM-based reasoning, verification-based re-ranking is applied only to the top-10 candidate sources returned by the selected similarity-based re-ranker. The verification stage is formulated as a listwise selection strategy, where the complete candidate set is considered jointly and the source that best supports the claim is selected. A structured zero-shot prompt is used to guide the model through several verification steps, including evidence consistency between claims and source abstracts, contextual specificity, and the exclusion of semantically similar but potentially conflicting sources. The prompt template is detailed in Figure~\ref{fig:verification_prompt} in Appendix. The selected source receives an additional score of +1 on top of the normalized ranking scores, ensuring that it is promoted to the first rank position while preserving the relative ordering of all remaining candidates. We consider three open-source generative LLMs as verification-based re-ranking models to analyze the effect of model architecture and model scale.

\bslabel{Gemma-4.} The \textit{google/gemma-4-31B-it} model is an instruction-tuned dense LLM with approximately 31B parameters from the Gemma family. Its reasoning capabilities and multilingual support make it suitable for verification-based re-ranking in multilingual scientific claim--source retrieval.

\bslabel{MiniMax-M2.5.} The \textit{MiniMaxAI/MiniMax-M2.5} model is a sparse mixture-of-experts (MoE) model with approximately 229B total parameters and around 10B activated parameters during inference. Its sparse activation enables efficient reasoning with substantially larger effective model capacity.

\bslabel{Kimi-K2.6.} The \textit{moonshotai/kimi-k2.6} model is a large-scale MoE model with approximately 1T total parameters and 32B activated parameters during inference. Its architecture provides higher capacity through sparse activation, enabling comparison against frontier-scale reasoning models.

\section{Evaluation}
\label{sec:evaluation}

We adopt a stage-wise evaluation strategy in which each component is evaluated sequentially and the strongest configuration is selected for subsequent stages. We first analyze different first-stage retrieval models across three claim representations, followed by the evaluation of similarity-based and verification-based re-ranking approaches. This setup incrementally refines the overall system configuration while enabling the contribution of each stage to be assessed individually. Following the CheckThat! 2026 setup, all experiments are evaluated using \textit{Mean Reciprocal Rank at 5} (MRR@5).

\subsection{First-Stage Retrieval}
\label{sec:eval_retrieval}

Table~\ref{tab:results_multiliguality} summarizes the first-stage retrieval results across different claim representations. Among the evaluated retrieval models, GritLM achieves the strongest overall performance, reaching the highest average MRR@5 of 0.632 with translated claims. For BM25, GTR, E5, and the original GritLM model, translated claims consistently yield the strongest results, indicating that reducing the language mismatch between non-English claims and predominantly English source publications improves retrieval. In contrast, the fine-tuned variant GritLM-F changes this behavior, reaching its highest performance with bilingual claims and improving German claims by +0.063 MRR@5 over the original GritLM model. While GritLM-F also improves French claims (+0.040), it substantially reduces performance for English claims (-0.122), showing that fine-tuning affects retrieval behavior differently across languages. One possible explanation is that fine-tuning on multilingual claim–source pairs and hard negatives adapts GritLM more strongly to cross-lingual retrieval, improving German and French claims but reducing the strong zero-shot English performance of the original model. To prioritize robust generalization to unseen claims, we adopt a conservative selection strategy and use the original GritLM model for English and French claims, while applying GritLM-F only for German claims where improvements are most pronounced and consistent across claim representations.

\begin{table}[t!]
\centering
\caption{First-stage retrieval performance across different claim representations for handling multilingual claims. We report MRR@5 for English (EN), German (DE), and French (FR) claims, while AVG denotes the arithmetic mean across all languages. GritLM-F denotes the fine-tuned variant of GritLM.}
\label{tab:results_multiliguality}
\setlength{\tabcolsep}{2pt}
\begin{tabular}{lcccc@{\hspace{6pt}}cccc@{\hspace{6pt}}cccc}
\toprule
\multirow{2}{*}{\begin{tabular}[c]{@{}l@{}}
\textbf{Retrieval}\\
\textbf{Models}
\end{tabular}} 
& \multicolumn{4}{c}{\textbf{Original Claims}} 
& \multicolumn{4}{c}{\textbf{Translated Claims}} 
& \multicolumn{4}{c}{\textbf{Bilingual Claims}} \\
\cmidrule(lr){2-5} \cmidrule(lr){6-9} \cmidrule(lr){10-13}
& EN & DE & FR & AVG 
& EN & DE & FR & AVG 
& EN & DE & FR & AVG \\
\midrule
BM25   & 0.4317 & 0.0719 & 0.0618 & 0.1885 & 0.4317 & 0.2844 & 0.4310 & 0.3824 & 0.4317 & 0.1936 & 0.2024 & 0.2759 \\
GTR    & 0.4821 & 0.2839 & 0.4040 & 0.3900 & 0.4821 & 0.3814 & 0.5085 & 0.4573 & 0.4821 & 0.3532 & 0.4932 & 0.4429 \\
E5     & 0.5333 & 0.2543 & 0.3882 & 0.3919 & 0.5333 & 0.4552 & 0.5688 & 0.5191 & 0.5333 & 0.4263 & 0.5319 & 0.4972 \\
GritLM & \underline {0.6595} & 0.5163 & 0.6567 & 0.6108 & 0.6595 & 0.5670 & 0.6707 & \underline {0.6324}  & 0.6595 & 0.5347 & 0.6612 & 0.6185 \\
\midrule
GritLM-F & 0.5380 & 0.5808 & 0.6631 & 0.5940 & 0.5380 & 0.5895 & 0.6799 & 0.6025 & 0.5380 & \underline {0.5974} & \underline {0.7011} & 0.6122 \\
\bottomrule
\end{tabular}
\end{table}

\subsection{Similarity-based Re-Ranking}
\label{sec:eval_similarity}

Similarity-based re-ranking substantially improves retrieval performance over the selected first-stage retrieval configuration, which combines GritLM for English and French claims with GritLM-F for German claims, as shown in Table~\ref{tab:results_similarity_reranking}. The strongest overall performance is achieved by Qwen3-8B-IT with the instruction-tuned prompt, improving average MRR@5 by +0.103 to 0.7419. Standard Qwen3-8B achieves comparable performance (+0.099), while Nemotron also provides notable improvements (+0.054). In contrast, Qwen3-0.6B slightly underperforms the baseline (-0.006), whereas BGE substantially reduces performance (-0.109). Interestingly, the customized Qwen3-8B prompt improves retrieval performance for English and French claims, while the standard prompt remains slightly stronger for German claims. Since the instruction-tuned version achieves the strongest overall retrieval performance, we select Qwen3-8B-IT for all languages in the final system configuration.

\begin{table}[t]
\centering
\caption{Similarity-based re-ranking performance on top of the selected first-stage retrieval configuration. We report MRR@5 for English (EN), German (DE), and French (FR) claims, while AVG denotes the arithmetic mean across all languages. Baseline denotes the selected first-stage retrieval configuration using GritLM for English and French claims and GritLM-F for German claims.}
\label{tab:results_similarity_reranking}
\setlength{\tabcolsep}{4pt}

\begin{tabular}{lcccc}
\toprule
\textbf{Re-ranker} & \textbf{EN} & \textbf{DE} & \textbf{FR} & \textbf{AVG} \\
\midrule
GritLM-(F) & 0.6595 & 0.5974 & 0.6612 & 0.6394 \\
\midrule
Nemotron & 0.7138 & 0.6285 & 0.7367 & 0.6930 \\
BGE & 0.5421 & 0.4866 & 0.5615 & 0.5301 \\
Jina & 0.6885 & 0.5910 & 0.7108 & 0.6634 \\
Qwen3-0.6B & 0.6527 & 0.5712 & 0.6750 & 0.6330 \\
Qwen3-8B & 0.7447 & \underline{0.6919} & 0.7773 & 0.7380 \\
\midrule
Qwen3-8B-IT & \underline{0.7536} & 0.6847 & \underline{0.7874} & \underline{0.7419} \\
\bottomrule
\end{tabular}
\end{table}

\subsection{Verification-based Re-Ranking}
\label{sec:eval_verification}

Verification-based re-ranking further improves retrieval performance on top of the selected first-stage retrieval and similarity-based re-ranking configuration (see Table~\ref{tab:results_signal_reranking}). Starting from the top-10 candidates generated by GritLM-(F) and Qwen3-8B-IT, the verification-based re-ranker selects the single candidate source that best supports the claim. All evaluated verification models improve retrieval performance over the similarity-based re-ranking baseline. Gemma-4 improves average MRR@5 by +0.013, while MiniMax-M2.5 achieves a slightly larger gain of +0.015. The strongest overall performance is achieved by Kimi-K2.6, increasing average MRR@5 from 0.7419 to 0.7648 (+0.023). Compared to earlier stages of the framework, the improvements introduced by verification-based re-ranking are smaller overall. However, the consistent gains across all evaluated models show that verification signals provide complementary information beyond similarity estimation alone.

\begin{table}[t]
\centering
\caption{Verification-based re-ranking performance on top of the selected first-stage retrieval and similarity-based re-ranking configuration. We report MRR@5 for English (EN), German (DE), and French (FR) claims, while AVG denotes the arithmetic mean across all languages. Baseline denotes the selected retrieval configuration using GritLM-(F) for first-stage retrieval with Qwen3-8B-IT for similarity-based re-ranking.}
\label{tab:results_signal_reranking}
\setlength{\tabcolsep}{4pt}
\begin{tabular}{lcccc}
\toprule
\textbf{Re-ranker} & \textbf{EN} & \textbf{DE} & \textbf{FR} & \textbf{AVG} \\
\midrule
GritLM-(F)+Qwen3-8B-IT        & 0.7536 & 0.6847 & 0.7874 & 0.7419 \\
\midrule
Verification (Gemma-4)   & 0.7662 & 0.6937 & 0.8043 & 0.7547 \\
Verification (MiniMax-M2.5)  & 0.7685 & 0.7001 & 0.8022 & 0.7569 \\
Verification (Kimi-K2.6)  & 0.7834 & 0.7176 & 0.7933 & 0.7648 \\
\bottomrule
\end{tabular}
\end{table}

\subsection{CheckThat! 2026 Leaderboard}
\label{sec:eval_sharedtask}

The final submitted system combines bilingual claim representations with metadata-enhanced source representations. First-stage retrieval uses \textit{GritLM} for English and French claims and the fine-tuned variant \textit{GritLM-F} for German claims. Candidate rankings are subsequently refined using the re-ranker \textit{Qwen3-8B-IT} for similarity-based re-ranking and \textit{Kimi-K2.6} for verification-based re-ranking.

Table~\ref{tab:official_results} reports the official test set results on CheckThat! 2026 Task 1 on \textit{Source Retrieval for Scientific Web Claims}. The submitted system achieved first place out of 37 participating teams, with an average MRR@5 score of 0.7628. Across individual languages, the system ranked first for English and French claims, and second for German claims. The strongest performance was achieved for French claims (0.7912), followed by English (0.7819) and German (0.7153).

\begin{table}[t]
\centering
\caption{Official test-set results of the submitted Claim2Source system on CheckThat! 2026 Task 1. We report MRR@5 for English (EN), German (DE), and French (FR) claims, while AVG denotes the arithmetic mean across all languages. Rk. denotes the rank achieved on the official leaderboard. Values in parentheses denote the score difference to the next-ranked system, where positive values indicate a higher score than the next lower-ranked system and negative values indicate a lower score than the next higher-ranked system.}
\label{tab:official_results}
\setlength{\tabcolsep}{2pt}
\begin{tabular}{lcccccccc}
\toprule
& \multicolumn{2}{c}{\textbf{EN}} 
& \multicolumn{2}{c}{\textbf{DE}}
& \multicolumn{2}{c}{\textbf{FR}}
& \multicolumn{2}{c}{\textbf{AVG}}\\
\cmidrule(lr){2-3}
\cmidrule(lr){4-5}
\cmidrule(lr){6-7}
\cmidrule(lr){8-9}
\textbf{System}
& \textbf{MRR@5} & \textbf{Rk.}
& \textbf{MRR@5} & \textbf{Rk.}
& \textbf{MRR@5} & \textbf{Rk.}
& \textbf{MRR@5} & \textbf{Rk.}\\
\midrule
Claim2Source
& \underline{0.7819} (+0.0017) & 1
& 0.7153 (-0.0075) & 2
& \underline{0.7912} (+0.0160) & 1
& \underline{0.7628} (+0.0048) & 1\\
\bottomrule
\end{tabular}
\end{table}

\section{Discussion}
\label{sec:discussion}

The experimental findings provide insights into the design decisions underlying the proposed system and highlight several remaining challenges for multilingual scientific claim--source retrieval.

\subsection{Design Choices Behind Claim2Source}
\label{sec:design_choices}

\bslabel{Structured Claim and Source Representations.} The results highlight that multilingual scientific claim--source retrieval benefits from structured claim and source representations. Bilingual claim representations combined with fine-tuned GritLM-F, in particular, improve performance for German (+0.063 MRR@5) and French (+0.040) claims. This indicates that preserving information from the original claim formulation provides complementary retrieval signals beyond translation alone. Additionally, metadata-enhanced source representations combine publication content with contextual information, such as authors and venue, to support the identification of indirect claim--source relationships throughout the retrieval and verification stages.

\bslabel{Progressive Candidate Refinement.} Retrieval performance increases progressively across multiple refinement stages. First-stage retrieval reduces the search space from 10,000 publications to a set of 100 candidate sources, while similarity-based re-ranking provides the largest performance gain (+0.103 MRR@5). Verification-based re-ranking further improves retrieval performance (+0.023) by refining only the top-10 candidate sources, although gains become smaller in later stages. Different retrieval stages contribute complementary signals, while early candidate reduction enables computationally expensive reasoning-based approaches to be applied to a small set of highly relevant candidates.

\subsection{Limitations and Future Work}
\label{sec:limitations}

\bslabel{Computational Costs.} The proposed framework relies on several computationally intensive components, including GritLM-7B for first-stage retrieval, Qwen3-8B for similarity-based re-ranking, and Kimi-K2.6 for verification-based re-ranking, which comprises approximately 1T total parameters with 32B active parameters during inference. Although smaller alternatives were evaluated throughout different stages of the framework, they consistently resulted in lower retrieval performance. This trade-off highlights an important challenge for multilingual scientific claim--source retrieval, namely improving computational efficiency without sacrificing retrieval effectiveness. Future work should investigate lightweight retrieval and re-ranking architectures or model distillation approaches that reduce computational costs while preserving retrieval performance.

\bslabel{Multilingual Generalization.} Despite its multilingual design, the framework remains sensitive to language-specific adaptations in terms of retrieval performance. While fine-tuned GritLM-F improves retrieval performance for German and French claims, it substantially reduces performance for English claims (-0.122 MRR@5). Consequently, the final system requires language-specific model selection and different retrieval configurations for each language rather than a unified multilingual setup. This underscores a significant challenge in the field of multilingual scientific claim--source retrieval: developing retrieval models that generalize robustly across languages while maintaining consistent performance. Although language-specific adaptation improves retrieval performance for German over the original GritLM model, German claims remain the most challenging. Understanding the remaining performance gap requires a more detailed error analysis and is left for future work.
\section{Conclusion}
\label{sec:conclusion}

In this paper, we presented a multi-stage retrieval framework for multilingual scientific claim--source retrieval in CheckThat! 2026 Task 1: \textit{Source Retrieval for Scientific Web Claims}. Our approach combines structured claim and source representations with progressive candidate refinement through first-stage retrieval, similarity-based re-ranking, and verification-based re-ranking. Results show that each stage contributes to retrieval performance, with smaller but complementary improvements across the framework. Our system achieves an average MRR@5 score of 0.7628 across English, German, and French claims, ranking first on the CheckThat! 2026 leaderboard. Future work should focus on improving multilingual generalization and reducing computational costs while preserving retrieval performance.

\begin{acknowledgments}
The authors acknowledge the financial support by the Federal Ministry of Research, Technology and Space of Germany (BMFTR) and by the Saxon State Ministry for Science, Culture and Tourism in the programme Center of Excellence for AI Research „Center for Scalable Data Analytics and Artificial Intelligence Dresden/Leipzig“, project identification number: ScaDS.AI. 

Tobias Schreieder is supported by the BMFTR through a Software Campus project, project identification number: 16|S23070.

The authors also acknowledge computing resources provided by the NHR Center at TU Dresden, supported by the BMFTR and the participating state governments within the NHR framework.
\end{acknowledgments}

\section*{Availability}
We publicly release the code for all experiments at \url{https://github.com/faerber-lab/CheckThat2026}.

\section*{Declaration on Generative AI}
The authors used ChatGPT for language editing and minor formatting assistance, and GitHub Copilot, Gemini, and GLM for coding assistance. These tools did not contribute to the intellectual content or scientific conclusions. After using these tools, all content was reviewed by the authors, who assume full responsibility for the publication.

\bibliography{literature}

\appendix

\section{GritLM Fine-Tuning Details}
\label{sec:appendix_finetuning}

This appendix provides the hyperparameters and training configuration used for LoRA fine-tuning of GritLM-F. The setup includes optimization settings, negative sampling configuration, and training infrastructure details referenced in Section~\ref{sec:methodology}.

\begin{table}[ht]
\centering
\caption{Hyperparameters and training configuration used for LoRA fine-tuning of GritLM-F. Training was performed on a single NVIDIA H100 GPU (see Section~\ref{sec:retrieval}).}
\label{tab:gritlm_finetuning}
\setlength{\tabcolsep}{6pt}
\begin{tabular}{lc}
\toprule
\textbf{Parameter} & \textbf{Value} \\
\midrule
Training epochs &  3\\
Learning rate &  2e\textsuperscript{-4}\\
Batch size & 8 \\
Gradient accumulation steps & 4 \\
Maximum sequence length & 256 \\
LoRA rank ($r$) & 16 \\
LoRA alpha & 64 \\
LoRA dropout &  0.1\\
Optimizer & AdamW \\
Learning rate scheduler & Cosine \\
Warmup ratio & 0.05 \\
Weight decay & 0.01 \\
Negative sampling strategy & Hybrid (BM25 + GritLM weighted fusion) \\
Number of negative samples & 8 \\
Number of hard negative samples & 6 \\
Number of random negative samples & 2 \\
\bottomrule
\end{tabular}
\end{table}

\section{LLM Prompt Templates}
\label{sec:appendix_prompts}

The following figures present the prompt templates used across the evaluated LLM-based retrieval and re-ranking components described in Section~\ref{sec:methodology}. The prompts include query translation, instruction-tuning for similarity-based re-ranking, and verification-based re-ranking.

\begin{figure}[ht]
\centering
\begin{tcolorbox}[
    colback=gray!5,
    colframe=gray!60,
    title={Translation Prompt},
    width=\textwidth,
    sharp corners,
    boxrule=0.5pt
]

\textbf{[Role \& Objective]}

You are a professional translator. Translate the following text to English.

\vspace{0.5em}

\textbf{[Instructions \& Constraints]}

Output ONLY the English translation — nothing else. Do not add explanations, notes, or quotes.

\end{tcolorbox}
\caption{Prompt template used for query translation with the  Qwen/Qwen3-8B model (see Section~\ref{sec:multilingual_claims}).}
\label{fig:translation_prompt}
\end{figure}

\begin{figure}[ht]
\centering
\begin{tcolorbox}[
    colback=gray!5,
    colframe=gray!60,
    title={Similarity-based Re-Ranking Instruction-Tuning Prompt},
    width=\textwidth,
    sharp corners,
    boxrule=0.5pt
]

\textbf{[Role \& Objective]}

You are a high-precision scientific fact-checker expert. Your mission is to determine whether a research paper (\texttt{DOCUMENT}) supports the claim made in a tweet (\texttt{QUERY}).

\vspace{0.5em}

\textbf{[Pre-Processing Rules]}

\begin{itemize}
\item Ignore emojis, slang, and non-scientific conversational fillers in the query.
\item Focus exclusively on the core scientific facts and claims.
\item Treat \texttt{\#} and \texttt{@} as noise unless they provide explicit metadata such as journal or author names.
\end{itemize}

\vspace{0.5em}

\textbf{[Analysis Dimensions]}

\begin{enumerate}
\item \textbf{Subject \& Entities:} Identify core research subjects and terminology. Check for overlap in proper nouns such as diseases, drugs, proteins, or methods.

\item \textbf{Methodology:} Compare the research frameworks or experimental approaches used in the paper and implied in the query.

\item \textbf{Results \& Data:} Prioritize matching specific numerical results, statistics, or scientific findings.

\item \textbf{Metadata Bonus (@Author):} Increase confidence if a name following \texttt{@} matches one of the paper authors.

\item \textbf{Metadata Bonus (\#Journal):} Increase confidence if a venue following \texttt{\#} matches the publication venue of the paper.
\end{enumerate}

\vspace{0.5em}

\textbf{[Output Instruction]}

\begin{itemize}
\item Respond with \texttt{yes} if the document provides sufficient evidence to support or strongly relate to the query.
\item Otherwise, respond with \texttt{no}.
\item Output strictly either \texttt{yes} or \texttt{no}. No explanations.
\end{itemize}

\end{tcolorbox}
\caption{Prompt template used for the Qwen3-8B-IT similarity-based re-ranker (see Section~\ref{sec:similarity_reranking}).}
\label{fig:qwen_reranker_prompt}
\end{figure}

\begin{figure}[ht]
\centering
\begin{tcolorbox}[
    colback=gray!5,
    colframe=gray!60,
    title={Verification-based Re-Ranking Prompt},
    width=\textwidth,
    sharp corners,
    boxrule=0.5pt
]

\textbf{[Role \& Objective]}

You are an expert research assistant. Your task is to identify the single most relevant paper from a provided list of candidates based on a specific user query.

\vspace{0.5em}

\textbf{Handling bilingual queries:}

The user may provide a single query (\texttt{QUERY: ...}) OR a pair (\texttt{QUERY (original language): ...} and \texttt{QUERY (translated to English): ...})

\begin{itemize}
\item If both are given, the original is in French or German. Use it as the PRIMARY source for matching.
\item The English translation is a SUPPLEMENT to help recognise cross-lingual synonyms and terminology.
\item Always consider both versions when evaluating relevance.
\end{itemize}

\vspace{0.5em}

\textbf{[Instructions \& Constraints]}

Analyze the Query and the Candidates below. Select the ONE paper that is the most relevant. Before making your final decision, you must perform the following checks:

\begin{itemize}
\item \textbf{Source Verification:} Does the paper title appear verbatim (or near-verbatim) in the query? (e.g., ``Ultrapotent antibodies...'')
    
\item \textbf{Evidence Alignment:} Does the paper's abstract explicitly support the specific claim made in the query? (e.g., If the query claims ``increased deaths,'' does the paper's results section confirm increased deaths?)
    
\item \textbf{Contextual Specificity:} Does the specific population, location, or intervention in the paper match the query? (e.g., ``US Veterans'' vs. ``France''; ``First Wave'' vs. ``Vaccinated'')
    
\item \textbf{Exclusion Check:} Are there other papers with similar keywords but contradictory conclusions? Discard them.
\end{itemize}

\vspace{0.5em}

You MUST respond with ONLY a JSON object with exactly two keys:

\begin{itemize}
\item \texttt{"reasoning"}: a brief string (2--4 sentences) explaining why this specific paper was chosen over the others based on the checks above. Mention any title matches or specific data points that aligned.

\item \texttt{"selected\_paper"}: an integer (1-based), the paper number that is most relevant.
\end{itemize}

\vspace{0.5em}

\textbf{Example response:}

\begin{lstlisting}
{
  "reasoning": "Paper 3 title matches the drug name in the query verbatim. Its abstract confirms the 50% reduction claim in the specific population mentioned. No other paper addresses this exact intervention.",
  "selected_paper": 3
}
\end{lstlisting}

No extra text, no explanation outside the JSON.

\end{tcolorbox}
\caption{Prompt template used for verification-based re-ranking across all evaluated LLMs (see Section~\ref{sec:verification_reranking}).}
\label{fig:verification_prompt}
\end{figure}

\end{document}